%% file: vnc-2014.tex
\documentclass[conference]{IEEEtran}

\ifCLASSINFOpdf
\else
\fi

\ifCLASSOPTIONcompsoc
  \usepackage[caption=false,font=normalsize,labelfont=sf,textfont=sf]{subfig}
\else
  \usepackage[caption=false,font=footnotesize]{subfig}
\fi
\usepackage[cmex10]{amsmath}

\ifCLASSINFOpdf
   \usepackage[pdftex]{graphicx}
\else
   \usepackage[dvips]{graphicx}
\fi
\usepackage{amssymb}

\usepackage{epstopdf}
\usepackage{graphicx}
\usepackage[caption=false]{subfig}

\usepackage{enumerate}
\usepackage{slashbox}

\usepackage{verbatim}

\usepackage{microtype}
\DisableLigatures{encoding = *, family = * }

\usepackage[nolist]{acronym}
\input{./acronym.tex}

\usepackage{makecell}

\begin{document}
\title{Towards Deploying a Scalable \& Robust Vehicular Identity and Credential Management Infrastructure}

\author{\IEEEauthorblockN{Mohammad Khodaei, Hongyu Jin, Panos Papadimitratos}
\IEEEauthorblockA{Networked Systems Security Group \\
KTH Royal Institute of Technology \\
Stockholm, Sweden \\
\emph{\{khodaei, hongyuj, papadim\}}@kth.se}}

\maketitle

\begin{abstract}
Several years of academic and industrial research efforts have converged to a common understanding on fundamental security building blocks for the upcoming \ac{VC} systems. There is a growing consensus towards deploying a \ac{VPKI} enables pseudonymous authentication, with standardization efforts in that direction. However, there are still significant technical issues that remain unresolved. Existing proposals for instantiating the \ac{VPKI} either need additional detailed specifications or enhanced security and privacy features. Equally important, there is limited experimental work that establishes the \ac{VPKI} efficiency and scalability. In this paper, we are concerned with exactly these issues. We leverage the common \ac{VPKI} approach and contribute an enhanced system with precisely defined, novel features that improve its resilience and the user privacy protection. In particular, we depart from the common assumption that the \ac{VPKI} entities are fully trusted and we improve user privacy in the face of an \emph{honest-but-curious} security infrastructure. Moreover, we fully implement our \ac{VPKI}, in a standard-compliant manner, and we perform an extensive evaluation. Along with stronger protection and richer functionality, our system achieves very significant performance improvement over prior systems - contributing the most advanced \ac{VPKI} towards deployment.

\end{abstract}
\IEEEpeerreviewmaketitle

\input{introduction}
\input{problem-statement}
\input{system-description}
\input{security-analysis}
\input{evaluation}
\input{related-works}

\input{conclusions}

\bibliographystyle{IEEEtran}
\bibliography{IEEEabrv,references}

\end{document}

%% file: acronym.tex
\begin{acronym}
	\acro{AAA}{Authentication, Authorization and Accounting}
	\acro{ACL}{Access Control List}
	\acro{AKI}{Accountable Key Infrastructure}
	\acro{API}{Application Programming Interface}
    \acro{BSM}{Basic Safety Message}
    \acro{BYOD}{Bring Your Own Device}
	\acro{C2C-CC}{Car2Car Communication Consortium}
	\acro{C2C}{Car-to-Car}
	\acro{C2I}{Car-to-Infrastructure}
	\acro{CA}{Certification Authority}
	\acro{CN}{Common Name}
	\acro{CAM}{Cooperative Awareness Message}
	\acro{CIA}{Confidentiality, Integrity and Availability}
	\acro{CRL}{Certificate Revocation List}
	\acro{CDN}{Content Delivery Network}
	\acro{COCA}{Cornell OnLine Certification Authority}
	\acro{CSR}{Certificate Signing Request}
	\acro{DAA}{Direct Anonymous Attestation}
	\acro{DDoS}{Distributed Denial of Service}
	\acro{DDH}{Decisional Diffie-Helman}
	\acro{DENM}{Decentralized Environmental Notification Message}
	\acro{DHT}{Distributed Hash Table}
	\acro{DoS}{Denial of Service}
	\acro{DPA}{Data Protection Agency}
	\acro{ECU}{Electronic Control Unit}
	\acro{ETSI}{European Telecommunications Standards Institute}
	\acro{ECDSA}{Elliptic Curve Digital Signature Algorithm}
	\acro{ECC}{Elliptic Curve Cryptography}
	\acro{EVITA}{E-safety Vehicle Intrusion protected Applications}
	\acro{FOT}{Field Operational Testing}
	\acro{FPGA}{Field-Programmable Gate Array}
	\acro{GPA}{Global Passive Adversary}
	\acro{GN}{GeoNetworking}
	\acro{GS}{Group Signatures}
	\acro{GM}{Group Manager}
	\acro{GBA}{Generic Bootstrapping Architecture}
	\acro{GUI}{Graphic User Interface}
	\acro{HSM}{Hardware Security Module}
	\acro{HTTP}{Hypertext Transfer Protocol}
	\acro{IEEE}{Institute of Electrical and Electronics Engineers}
	\acro{IETF}{Internet Engineering Task Force}
	\acro{IoT}{Internet of Things}
	\acro{ITS}{Intelligent Transport Systems}
	\acro{IT}{Information Technologies}
	\acro{IMSI}{International Mobile Subscriber Identity}
	\acro{IMEI}{International Mobile Station Equipment Identity}
	\acro{IdP}{Identity Provider}
	\acro{IDS}{Intrusion Detection System}
	\acro{ISP}{Internet Service Provider}
	\acro{LEA}{Law Enforcement Agency}
	\acro{LTC}{Long Term Certificate}
	\acro{LTCA}{Long Term \acl{CA}}
	\acro{LDAP}{Lightweight Directory Access Protocol}
	\acro{LTE}{Long Term Evolution}
	\acro{LBS}{Location Based Service}
	\acro{MAC}{Message Authentication Code}
	\acro{MCA}{Message \ac{CA}}
	\acro{MEA}{Misbehavior Evaluation Authority}
	\acro{OBU}{On-Board Unit}
	\acro{OCSP}{Online Certificate Status Protocol}
	\acro{PCA}{Pseudonym \acl{CA}}
	\acro{PDP}{Policy Decision Point}
	\acro{PEP}{Policy Enforcement Point}
	\acro{PIR}{Private Information Retrieval}
	\acro{PKI}{Public-Key Infrastructure}
	\acro{PRECIOSA}{Privacy Enabled Capability in Co-operative Systems and Safety Applications}
	\acro{PRESERVE}{Preparing Secure Vehicle-to-X Communication Systems}
	\acro{P2P}{peer-to-peer}
	\acro{PS}{Participatory Sensing}
	\acro{RA}{Resolution Authority}
	\acro{REST}{Representational State Transfer}
	\acro{RBAC}{Role Based Access Control}
	\acro{RCA}{Root Certification Authority}

	\acro{RSU}{Roadside Unit}
	\acro{SAML}{Security Assertion Markup Language}
	\acro{SAS}{Sample Aggregation Service}
	\acro{SCMS}{Security Credential Management System}
	\acro{SCORE@F}{Système COopératif Routier Expérimental Français}
	\acro{SDSI}{Simple Distributed Security Infrastructure}
	\acro{SeVeCom}{Secure Vehicle Communication}
	\acro{SIT}{Sichere Informationstechnologie}
	\acro{SoA}{Service-oriented-Approach}
	\acro{SSO}{Single-Sign-On}
	\acro{SSL}{Secure Sockets Layer}
	\acro{SOAP}{Simple Object Access Protocol}
	\acro{TACK}{Temporary Anonymous Certified Key}
	\acro{TS}{Task Service}
	\acro{TLS}{Transport Layer Security}
	\acro{TPM}{Trusted Platform Module}
	\acro{TVR}{Ticket Validation Repository}
	\acro{URI}{Uniform Resource Identifier}
	\acro{VANET}{Vehicular Ad-hoc Network}
	\acro{V2I}{Vehicle-to-Infrastructure}
	\acro{V2V}{Vehicle-to-Vehicle}
	\acro{V2X}{\ac{V2V} and/or \ac{V2I}}
	\acro{VC}{Vehicular Communication}
	\acro{VM}{Virtual Machine}
	\acro{VSS}{\ac{VC} Security Subsystem}
	\acro{WAVE}{Wireless Access in Vehicular Environments}
	\acro{WSDL}{Web Services Discovery Language}
	\acro{W3C}{World Wide Web Consortium}
	\acro{V}{Vehicle}
	\acro{VANET}{Vehicular Ad-hoc Network}
	\acro{VPKI}{Vehicular Public-Key Infrastructure}
	\acro{WS}{Web Service}
	\acro{WoT}{Web of Trust}
	\acro{WSACA}{\ac{WAVE} Service Advertisement \ac{CA}}
	\acro{XML}{Extensible Markup Language}
	\acro{XACML}{eXtensible Access Control Markup Language}
	\acro{3G}{3rd Generation}
\end{acronym}

%% file: introduction.tex
\section{Introduction}
\label{sec:introduction}

\acf{VC} systems enable a multitude of applications, disseminating warnings on environmental hazards, traffic conditions and other location-relevant information \cite{papadimitratos2009vehicular,ETSI-102-638}. At the same time, the need to secure \ac{VC} systems and protect their users' privacy has been long and well understood \cite{papadimitratos2006securing}, and academia and industry have worked towards addressing these challenges. The common understanding is to use public key cryptography: long-term credentials (and keys) for accountable identification of the \ac{VC} on-board platform (termed the vehicle for simplicity), in conjunction with short-term anonymized credentials, broadly termed \emph{pseudonyms}, for limited unlinkability of vehicle-originating messages. Projects on both sides of the Atlantic developed systems based on this approach (\ac{SeVeCom} \cite{papadimitratos2008secure}, CAMP \cite{whyte2013security}, IEEE 1609.2 WG \cite{1609draft}), in cooperation with harmonization (Car-to-Car Communication Consortium \cite{c2c,bissmeyer2011generic}) and standardization (\ac{ETSI} \cite{ETSI-102-731,ETSI-102-941}) bodies.

The cornerstone of all these efforts, along with continuing work towards \ac{FOT}, is a \acf{VPKI} that comprises a set of \emph{authorities} with distinct roles: the \ac{LTCA}, the \ac{PCA}, and the \ac{RA}. The \ac{LTCA} is responsible for issuing \acp{LTC}, in principle one per vehicle. The \ac{PCA} issues sets of pseudonyms to each vehicle registered with an \ac{LTCA}. Both the \ac{LTCA} and the \ac{PCA} can revoke the credentials they issued. When necessary, e.g., for investigation purposes, the \ac{RA} can initiate a process to reveal the long-term identity of a vehicle, based on a set of pseudonymously authenticated messages. This separation of duty provides conditional anonymity, revoked under special circumstances, while ensuring that only legitimate vehicles can obtain pseudonyms and accountably participate in the system.

\textbf{Challenges and Contributions:} With basic concepts understood, there are few works that crisply define \ac{VPKI} components. On that front, we advance the state-of-the-art, enhancing our earlier work for a \emph{multi-domain} \ac{VPKI} \cite{vespa, alexiou2013dspan}, with a more complete system. Our protocols presented here and their novel features render the \ac{VPKI} more robust to misbehaving vehicles. In particular, even in a future environment with a multiplicity of \ac{LTCA} and \ac{PCA} servers, it is impossible for a compromised vehicle to obtain multiple credentials valid simultaneously (i.e., set the ground for Sybil-based \cite{douceur2002sybil} misbehavior), and thus harm the \ac{VC} operations. Moreover, we propose a generic pseudonym lifetime determination approach to enhance message unlinkability, thus user privacy.

So far, it has been assumed, often implicitly, that the \ac{VPKI} servers are fully trustworthy. Nonetheless, the prospect of having multiple such servers commercially deployed, in diverse environments under different regulations, makes this assumption less realistic. In fact, one cannot preclude servers that are \emph{honest}, i.e., follow specified protocols and protect their private keys, but they may be \emph{curious}, i.e., tempted to trace clients (vehicles) if given the opportunity. For example, to offer customized services or optimize own operations. The experience from other mobile applications and location-based services hints this is a realistic threat to user privacy. To address this challenge, we extend our adversary model by considering \emph{honest-but-curious} servers and design our \ac{VPKI} to be resilient against such behaviors.

Last but not least, very few works provided detailed experimental validation of their \ac{VPKI} designs to show the performance and availability of their systems. To address this challenge, we develop a \emph{standard-compliant} full-fledged, refined, cross-platform \ac{VPKI} and present an extensive experimental evaluation. Using the similar setup as in the literature, to have a meaningful and direct comparison, we find that our system achieves very significant improvement over prior art. With contributions on these three dimensions, we advance towards a more robust and scalable concrete \ac{VPKI} system. 

In the rest of the paper, we describe the system and adversarial model considered (Sec. \ref{sec:problem-statement}) and move on with the design of our system (Sec. \ref{sec:solution}). We then analyze the protocols (Sec. \ref{sec:security-analysis}) and present extensive experimental evaluation (Sec. \ref{sec:evaluation}), before related work (Sec. \ref{sec:related-work}) and conclusions (Sec. \ref{sec:conclusions}). 

%% file: problem-statement.tex
\section{System \& Adversarial Model and Objectives}
\label{sec:problem-statement}

\textbf{System model \& assumptions:} We assume a \ac{VPKI} architecture with distinct entities (\ac{LTCA}, \ac{PCA}, and \ac{RA}), and we define a \emph{domain} as the set of vehicles registered with one \ac{LTCA}, subject to the same administrative regulations and policies. We do not dwell on the formation of such domains (geographic regions, cities, states, or otherwise). We assume that the \ac{LTCA} is reachable by vehicles registered with it. Furthermore, we assume multiple \ac{PCA} servers, active in one or across multiple domains, which have already established trust (security associations) with the corresponding \ac{LTCA}(s). Each vehicle has a unique membership, registered to one domain, its \emph{home domain}; it can freely obtain pseudonyms from any \ac{PCA} in its home or in a \emph{foreign} domain.\footnote{In the context of \ac{VC} systems, the notion of a foreign certificate was first introduced in \cite{papadimitratos2008certificate}.} We assume that across different domains, trust is established with the help of a higher-level authority, \ac{RCA}, or a set of such authorities and cross-certification. 

\textbf{Adversary model:} We adhere to the assumed adversarial behavior defined in the literature \cite{papadimitratos2006securing} and we are primarily concerned with adversaries that seek to abuse the \ac{VPKI}. With the multi-domain, thus multi-\ac{PCA}, environment, internal adversaries raise a specific challenge: they could seek to obtain multiple pseudonyms valid simultaneously over the same period of time. This would allow them to act as multiple legitimate vehicles at the same time, e.g., injecting multiple bogus messages and possibly control the outcome of specific protocols, e.g., involving voting \cite{raya2007eviction}. In addition, we care about external adversaries that may mount a clogging \ac{DoS} attack against the \ac{VPKI} servers. 

We are further concerned with \ac{VPKI} servers that are \emph{honest-but-curious}. A \ac{VPKI} server could have access to eavesdropped vehicle communications, that is transcripts of anonymized signed messages. Then, with knowledge obtained from the \ac{VPKI} operations, if it is able to link pseudonym sets provided to the same vehicle, it could create traces of vehicle activities, thus perform a sort of user profiling. In the worst case, multiple servers could \emph{collude}, i.e., share knowledge. 

\textbf{Objectives:} We seek to improve the protection achieved by strengthening the robustness of the \ac{VPKI} to adversarial attacks, notably in the light of a multi-domain setup. Moreover, we seek to improve the \ac{VPKI} in rendering it more resilient to \emph{honest-but-curious} servers. The motivation for the latter stems from experience in other areas of mobile computing: service providers tend to amass information in an attempt to profile clients. Although recent \ac{VPKI} proposals separate duties among servers \cite{vespa, gisdakis2013serosa}, no design explicitly sought to prevent such tracking. We wish to maintain standard-compliant functionalities, but at the same time protect privacy. Last but not least, we wish to significantly improve the efficiency of the \ac{VPKI} demonstrated through detailed experimental evaluations, towards scalable \ac{VC} system deployment. 

%% file: system-description.tex
\section{Our Solution}
\label{sec:solution}

\subsection{System Overview}
\label{subsec:system-overview}

%#######################################
\begin{figure}[t!]
    \centering
	\includegraphics[width=0.5\textwidth,height=0.5\textheight,keepaspectratio] {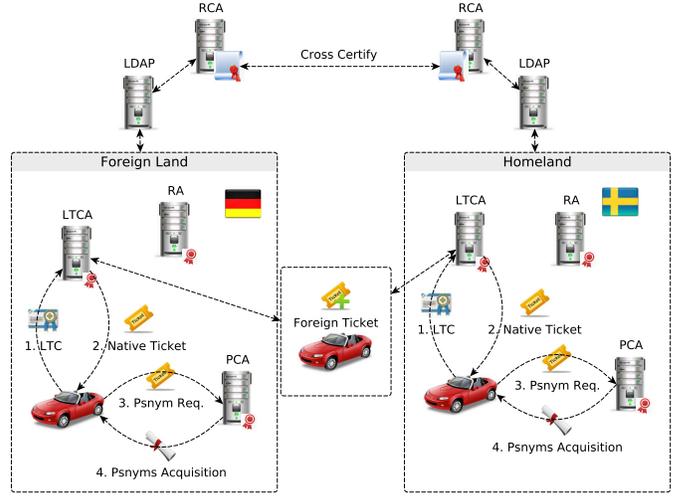} % picture file name
	\caption{System Overview}
	\label{fig:system-overview}
	\vspace{-1em}
\end{figure}
%#######################################

Fig. \ref{fig:system-overview} illustrates our proposed \ac{VPKI} with two domains. The \ac{LTCA} \emph{registers} vehicles and maintains their long-term identities. It then authenticates registered vehicles and grants them access to credential management services, prominently to \emph{obtain pseudonyms}. To do so, the vehicle obtains a \emph{native ticket} from its home \ac{LTCA} (H-\ac{LTCA}) and it presents it to any \ac{PCA} of its choice (e.g., one easily accessible, available, mandated, or simply preferred). The tickets are anonymized, in order not to reveal the vehicle identity to the \ac{PCA}. At the same time, the \emph{ticket issuance} protocol does not reveal to the \ac{LTCA} the targeted \ac{PCA} or the actual pseudonym acquisition period.  

If the vehicle moves to another domain, say from Sweden to Germany, it can request from its home \ac{LTCA} to obtain a \emph{foreign ticket} without revealing the targeted foreign \ac{LTCA} (F-\ac{LTCA}). It then presents the foreign ticket to the German \ac{LTCA}, to obtain a ticket to present to any associated \ac{PCA}.\footnote{The notions of native and foreign tickets are transparent to the home \ac{LTCA} based on the protocol design.} This way, the \ac{PCA} in the German domain (in our example) will not be able to classify its requester separately from other \emph{local} German vehicles. This is further analyzed in Sec. \ref{sec:security-analysis}.

The vehicle interacts with any \ac{PCA} to obtain new pseudonyms, fetch \acp{CRL} \cite{papadimitratos2008certificate, solo2002internet}, or validate pseudonym revocation status using the \ac{OCSP} \cite{myers1999x}. \ac{OCSP} requests are authenticated with a current valid pseudonym. The overall multi-domain operations (e.g., ticket and pseudonym acquisition in a foreign domain) are assisted by directory services (\ac{LDAP}). In case of misbehavior (e.g., detected locally by vehicles \cite{moore2008fast} or for other reasons \cite{Papadi:C:08}), the \ac{RA} is able to resolve a pseudonym and possibly revoke the pseudonyms and the \ac{LTC} of the misbehaving vehicle. We assume that the certificates of \acp{RCA} are pre-installed to \acp{OBU} (referred to as the vehicle for simplicity) and that the \ac{VPKI} servers and the vehicles are \emph{loosely synchronized}. All protocols are run over \ac{TLS}, with mutual authentication for obtaining tickets, and with unidirectional (server-only) authentication for obtaining pseudonyms and for \ac{LDAP} queries. Table \ref{table:protocols-notation} summarizes notation used in the constituent protocols. Within a domain, the issuance of pseudonyms is regulated by the same policy, further explained in Sec. \ref{sec:security-analysis}.

\subsection{\ac{VPKI} Services and Protocols}
\label{subsec:vpki-components}

\begin{figure}[t!]
    \centering
    \includegraphics[trim=5.4cm 11cm 5.35cm 11cm, clip=true, totalheight=0.18\textheight,angle=0] {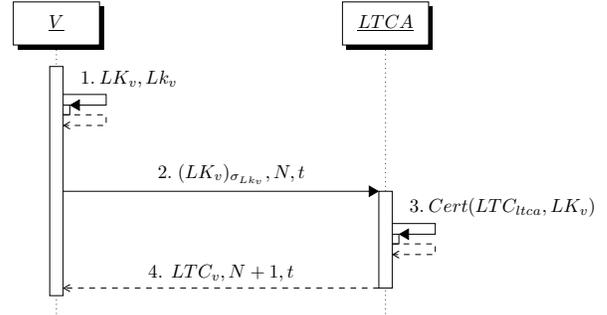} % picture file name
    \caption{Vehicle Registration and \ac{LTC} Update}
    \label{fig:vehicle-registration-and-certificate-update-process}
\end{figure}

\textbf{Vehicle Registration and \ac{LTC} Update (Fig. \ref{fig:vehicle-registration-and-certificate-update-process}):} Each vehicle is registered to its home \ac{LTCA} and it is issued an \ac{LTC}. The vehicle generates a pair of public and private keys, $LK_v$ and $Lk_v$. The prepared \ac{CSR} \cite{cooper2008internet} is sent to the home \ac{LTCA}.

%######################################
\begin{table}
	\caption{Notation used in the protocols}
	\centering
	\resizebox{0.45\textwidth}{!}
		{
		\renewcommand{\arraystretch}{1.13}
		\begin{tabular}{l | *{1}{c} r}
			\hline \hline
			$Lk$ & \emph{Long-term Private Key} \\\hline
			$LK$ & \emph{Long-term Public Key} \\\hline
			$k^i_v$ & \emph{Pseudonymous Private Key} \\\hline
			$K^i_v$ & \emph{Pseudonymous Public Key} \\\hline
			$P^i_v$ & \emph{Pseudonymous Certificate} \\\hline
			$CA_{id}$  & \emph{\acl{CA} Unique Identifier} \\\hline
			$LTC$  & \emph{\acl{LTC}} \\\hline
			$Cert(LTC, msg)$ & \emph{Processing of Signing on msg} \\\hline
			$\sigma_{CA}$ & \emph{Signature of \acl{CA}} \\\hline
			$t$ & \emph{Timestamp} \\\hline
			$N$ & \emph{Nonce} \\\hline
			$t_s$ & \emph{Starting Timestamp} \\\hline
			$t_e$ & \emph{Ending Timestamp} \\\hline
			$SN$ & \emph{Serial Number} \\\hline
			$H()$ & \emph{Hash Function} \\\hline
			$tkt_{expiry}$ & \emph{Ticket Expiration Time} \\\hline
			\hline
		\end{tabular}
		\renewcommand{\arraystretch}{1}
		\label{table:protocols-notation}
		}
\end{table}
%######################################

\begin{figure}[t!]
    \centering
    \includegraphics[trim=4.92cm 10.5cm 3.4cm 10.5cm, clip=true, totalheight=0.18\textheight,angle=0] {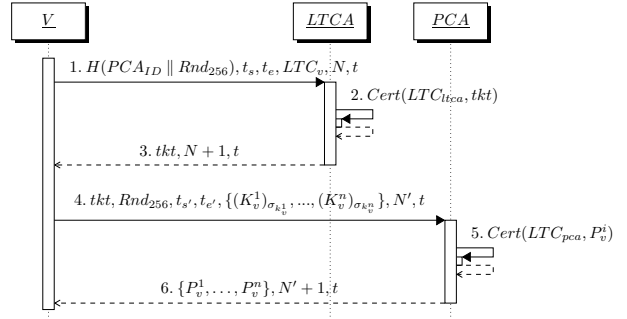} % picture file name
    \caption{Ticket and Pseudonym Acquisitions}
    \label{fig:ticket-provision-service-acquisition-in-direct-mode-diagram}
	\vspace{-1em}
\end{figure}

\textbf{Pseudonym Provision (Fig. \ref{fig:ticket-provision-service-acquisition-in-direct-mode-diagram}):} The vehicle, \emph{V}, calculates the hash value of the concatenation of the desired $PCA_{id}$ and a 256-bits random number; it chooses the desired time interval, [$t_s$, $t_e$] for which it will request pseudonyms, and appends its $\ac{LTC}_v$ to the \emph{ticket provisioning request}. The protocol to obtain a ticket is:

\begin{equation*} \label{msg:tkt-req-res}
	\begin{split}
	     \ac{V} \longrightarrow LTCA:& \: H(PCA_{id}\|Rnd_{256}), t_s, t_e, \ac{LTC}_v, N, t \\
	     \ac{LTCA} \longrightarrow V:& \: tkt, N+1, t
	\end{split}
\end{equation*}

The format of a ticket, signed by the \ac{LTCA}, is:
\vspace{-1px}
\begin{equation*} \label{msg:tkt-format}
	\begin{split}
		tkt = \{ & SN, H(PCA_{id} \| Rnd_{256}), t_s, t_e, tkt_{expiry}\}_{\sigma_{\text{LTCA}}}
	\end{split}
\end{equation*}

%#######################################
%#######################################

With the ticket obtained, the vehicle initiates the protocol to obtain pseudonyms:

\begin{equation*} \label{msg:psnym-req-res}
	\begin{split}
		 \ac{V} \longrightarrow PCA:& \: Rnd_{256}, t'_{s}, t'_{e}, tkt, \{(K^1_v)_{\sigma_{k^1_v}}, ..., (K^n_v)_{\sigma_{k^n_v}}\}, N', t \\
		 \ac{PCA} \longrightarrow V:& \: \{P^1_v, \dots, P^n_v\}, N'+1, t
	\end{split}
\end{equation*}

The \ac{PCA} verifies the hash value in the ticket by hashing the concatenation of its own identity and the provided random number. This ensures the ticket was issued for this \ac{PCA} for the exact said time interval. A \ac{CSR}, $(K^i_v)_{\sigma_{k^i_v}}$, is the signed public key generated by the vehicle. The period of requested pseudonyms, $ t'_{s}$ and $t'_{e}$, fall within the period of the ticket, $[t_s, \: t_e]$. Each pseudonym, signed by the \ac{PCA}, is: %shown below: 
\begin{equation*} \label{msg:psnym-format}
	P^i_v = \{SN, K^i_v, [t_{s}^i, t_{e}^i]\}_{\sigma_{\ac{PCA}}}
\end{equation*}

%#######################################
%#######################################

Each \ac{CSR} includes the proof of possession of the corresponding private key, $k^i_v$ \cite{myers1999internet}. Otherwise, the \ac{PCA} does not issue pseudonyms for the requester. If the proof of possession on a $K^i_v$ fails, the \ac{PCA} assumes a fault; it issues pseudonyms for the correctly signed $K^i_v$ and replies with an appropriate error message for the invalid signature. However, if the number of invalid proofs of possessions reaches a threshold, the \ac{PCA} deems the requester malicious and aborts. 

\begin{figure}[t!]
    \centering
    \includegraphics[trim=4.9cm 10.5cm 4.6cm 10.5cm, clip=true, totalheight=0.19\textheight,angle=0] {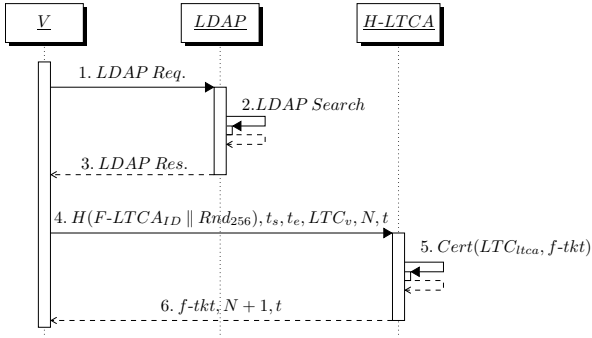} % picture file name
    \caption{Roaming User: Foreign Ticket Authentication}
    \label{fig:multi-domain-user-authentication-diagram}
\end{figure}

\begin{figure}[t!]
  \begin{center}
    \centering
    \includegraphics[trim=4.9cm 10.5cm 3.5cm 10.5cm, clip=true, totalheight=0.19\textheight,angle=0] {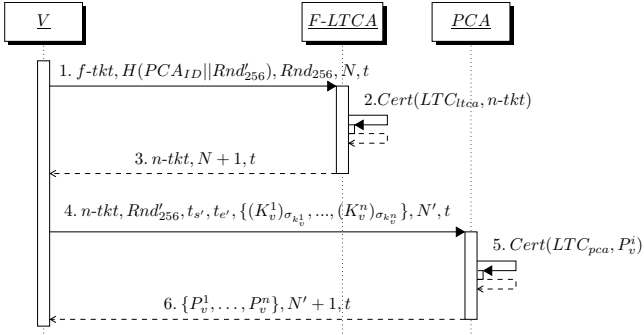} % picture file name
    \caption{Native Ticket and Pseudonym Acquisition in the Foreign Domain}
    \label{fig:multi-domain-user-authentication-service-acquisition-diagram}
	\vspace{-1em}
  \end{center}
\end{figure}

\textbf{Foreign Domain Pseudonym Acquisition (Fig. \ref{fig:multi-domain-user-authentication-diagram} and \ref{fig:multi-domain-user-authentication-service-acquisition-diagram}):} A vehicle crossing into a foreign domain obtains a \emph{foreign ticket} (\emph{f-tkt}) issued by its home \ac{LTCA} without disclosing the targeted domain. To obtain an \emph{f-tkt} from $H\textnormal{-}\ac{LTCA}$:

\begin{equation*} \label{msg:f-tkt-req-res}
	\begin{split}
	     \ac{V} \longrightarrow H\textnormal{-}\ac{LTCA}:& H(F\textnormal{-}LTCA_{id}\|Rnd_{256}), t_s, t_e, \ac{LTC}_v, N, t \\
	     H\textnormal{-}\ac{LTCA} \longrightarrow \ac{V}:& f\textnormal{-}tkt, N+1, t
	\end{split}
\end{equation*}

To obtain a \emph{native ticket} (\emph{n-tkt}) in the foreign domain, the vehicle sends its foreign ticket (instead of its $LTC_v$) to the \ac{LTCA} in the foreign domain, after finding its certificate with the help of an \ac{LDAP} server (similar to Fig. \ref{fig:ticket-provision-service-acquisition-in-direct-mode-diagram}).

\textbf{Pseudonym Resolution \& Revocation (Fig. \ref{fig:pseudonym-resolution-revocation-diagram}):} The \acf{RA}, in case of misbehavior, can resolve a pseudonym with the help of the \ac{PCA} and the \ac{LTCA}. The \ac{RA} first asks the \ac{PCA} to map the transcript pseudonyms to the corresponding ticket (which the \ac{PCA} returns). Then, the \ac{RA} queries the \ac{LTCA} to have the vehicle identified (as the \ac{LTCA} can map the ticket to the long-term credentials of the vehicle). In case of a cross-domain resolution, one additional step, involving the foreign \ac{LTCA} makes the link to the H-\ac{LTCA}, is needed. Steps 1, 2 and 3 in Fig. \ref{fig:pseudonym-resolution-revocation-diagram} show the process to resolve a ticket and possibly revoke the pseudonyms. The \ac{PCA} adds all the valid pseudonyms, issued for the ticket, to the \ac{CRL}. Steps 4, 5 and 6 show the process to identify the vehicle and revoke its \ac{LTC}.

%#######################################
%#######################################

%#######################################
%#######################################

%% file: security-analysis.tex
\section{Security Analysis}
\label{sec:security-analysis}

%#######################################
\begin{figure}[t!]
    \centering
    \includegraphics[trim=5cm 10.5cm 5.1cm 10.5cm, clip=true, totalheight=0.21\textheight,angle=0] {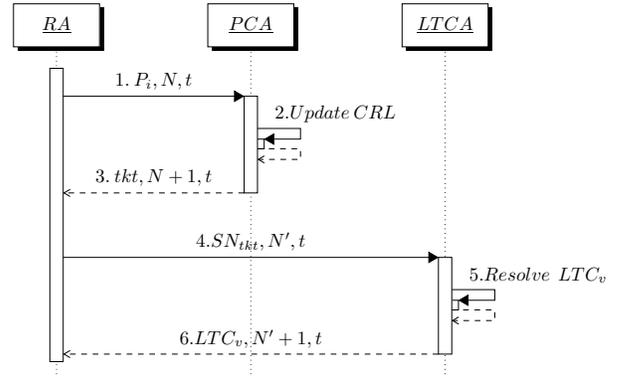} % picture file name
    \caption{Pseudonym Resolution and Revocation}
    \label{fig:pseudonym-resolution-revocation-diagram}
\end{figure}
%#######################################

\begin{table}
	\caption{Notation used in security analysis}
	\centering
	\resizebox{0.45\textwidth}{!}
		{
		\begin{tabular}{l | *{1}{c} r}
			\hline \hline
			$LTCA_{A}$ & \emph{\ac{LTCA} of domain A} \\\hline
			$PCA_{A_i}$ & \emph{$PCA_i$ in domain A} \\\hline
			$PCA_{A}$ & \emph{A set of \acp{PCA} in domain A} \\\hline
			$id_{A}$ & \emph{Identities of the vehicles in domain A} \\\hline
			$P_{A}$ & \emph{Pseudonyms issued by the \acp{PCA} in domain A} \\\hline
			\hline
		\end{tabular}
		\label{table:sec-analysis-notations}
		}
\end{table}
%######################################

Our primary concern is to analyze the achieved security and privacy considering the honest-but-curious \ac{VPKI} servers. We present these findings in Sec.~\ref{sec:security-analysis-formal}. Then, in Sec.~\ref{sec:security-analysis-pnym}, we explain how arbitrary pseudonym request times along with pseudonym lifetimes can provide significant information to any adversary (eavesdropper) to link pseudonyms and thus messages. Accordingly, we explain how our policy, enforced at \acp{PCA}, can mitigate this threat. Table \ref{table:sec-analysis-notations} summarizes notation used in the security analysis.

\textbf{Communication integrity, confidentiality, non-repudiation:} This is achieved thanks to the use of secure channels (\ac{TLS}) for vehicle to \ac{VPKI} communication, while security associations allow any server to authenticate messages generated by any other server (notably, tickets). Digital signatures and certificates ensure non-repudiation. 

%#######################################

\begin{table*}
	\caption{Information held by honest-but-curious Entities}
	\centering
	\Large
	\resizebox{1\textwidth}{!}
		{
		\renewcommand{\arraystretch}{1.8}
			\begin{tabular}{ | c | *{1}{c} | *{1}{c} | }
				\hline
				\Huge \textbf{Honest-but-curious (colluding) Entities} & \Huge \textbf{Information} & \Huge \textbf{Privacy Implications}\\\hline
	
			   	\Huge $LTCA_{A}$ & \Huge $id_A, t_s, t_e$ & \Huge An \ac{LTCA} knows during when the registered vehicles wish to obtain pseudonyms.\\\hline
	
			    \Huge $PCA_{A_i}$ & \Huge $t_s, t_e, P_{A_i}$ & \Huge A \ac{PCA} can link the pseudonyms it issued for a same request, but cannot link those for different requests.\\\hline
	
			    \Huge $LTCA_{A}$, $PCA_{A}$ & \Huge $id_A, t_s, t_e, P_{A}$ & \Huge The pseudonyms they issued can be linked and the vehicle identities within the same domain can be derived.\\\hline
	
			    \Huge $LTCA_{A}$, $LTCA_{B}$ & \Huge $id_A, id_B, t_s, t_e$ & \Huge Collusion among \acp{LTCA} from different domains does not reveal additional information.\\\hline
	
			    \Huge $PCA_{A}$, $PCA_{B}$ & \Huge $t_s, t_e, P_{A}, P_{B}$ & \Huge Collusion among \acp{PCA} from different domains does not reveal additional information.\\\hline
	
			    \Huge $LTCA_{A}$, \Huge $LTCA_{B}$, \Huge $PCA_{A}$, \Huge $PCA_{B}$ & \Huge $id_A, id_B, t_s, t_e, P_{A}, P_{B}$ & \Huge 
			    \makecell{Colluding \acp{LTCA} and \acp{PCA} can link the pseudonyms they issued. The vehicle identities can be derived, \\as long as the issuers of the pseudonyms and the corresponding tickets collude.}\\\hline
			\end{tabular}
			\renewcommand{\arraystretch}{1}
			\label{table:VPKI-entity-knowledge}
		}
\end{table*}

\textbf{Authentication and authorization:} The \ac{LTCA} makes the appropriate decisions, based on the registration of the vehicle, its status (revoked or not), and the use of its long-term credentials. The \ac{PCA} grants the service, the pseudonyms, by validating the \ac{LTCA} signature, based on their prior trust establishment. Trust associations of \acp{PCA} and \acp{LTCA} are made known to vehicles through \ac{LDAP} services.

\textbf{Thwarting Sybil-based misbehavior:} The \ac{LTCA} keeps the records of the issued tickets. Upon receiving a request, the \ac{LTCA} checks whether a ticket was issued to the vehicle for that period. This ensures that no vehicle can request more than one ticket for any period. As a ticket is bound to a specific \ac{PCA} and the \ac{PCA} keeps records of ticket usage, a ticket cannot be reused for other \acp{PCA}. This implies that at any point in time, the vehicle cannot obtain more than one pseudonym sets (thus pseudonyms) valid simultaneously. This means that one cannot act as more than one entity. 

\textbf{Concealing pseudonym providers, foreign identity providers and actual pseudonym acquisition period:} By sending $H(PCA_{id} \| Rnd_{256})$, $t_s$, $t_e$, $LTC_v$ to the \ac{LTCA}, the vehicle reveals its long-term identity, $LTC_v$, but it hides the targeted $\ac{PCA}_{id}$ (and the targeted $\ac{LTCA}_{id}$ in case of foreign ticket request) and the actual requested interval, $[t'_{s}, t'_{e}]$. The \ac{PCA}, then, verifies if $[t'_{s}, \: t'_{e}] \subseteq [t_s, \: t_e]$ and grants pseudonyms accordingly.

%#######################################

\subsection{Honest-but-curious Authorities}\label{sec:security-analysis-formal}

Table \ref{table:VPKI-entity-knowledge} outlines the information held by honest-but-curious, possibly colluding, authorities within the same domain or across domains. We do not consider the \ac{RA} because it does not hold information on unresolved pseudonyms. Of course, a malicious (but, not honest-but-curious) \ac{RA} could unnecessarily initiate resolutions. Collusion among the \ac{LTCA} and the \acp{PCA} from the same domain enables them to link vehicle identities to issued pseudonyms. Moreover, collusion among \acp{LTCA} and \acp{PCA} from different domains further enables them to link vehicle identities to pseudonyms issued for foreign tickets. 

\subsection{Tickets and Pseudonym Lifetime Policies}\label{sec:security-analysis-pnym}

\begin{figure}
    \centering
      \subfloat[Flexible lifetimes]{ %Non-overlapping Psnyms
		\includegraphics[trim=1.5cm 0.1cm 0.7cm 0.65cm, clip=true, totalheight=0.148\textheight,angle=0]{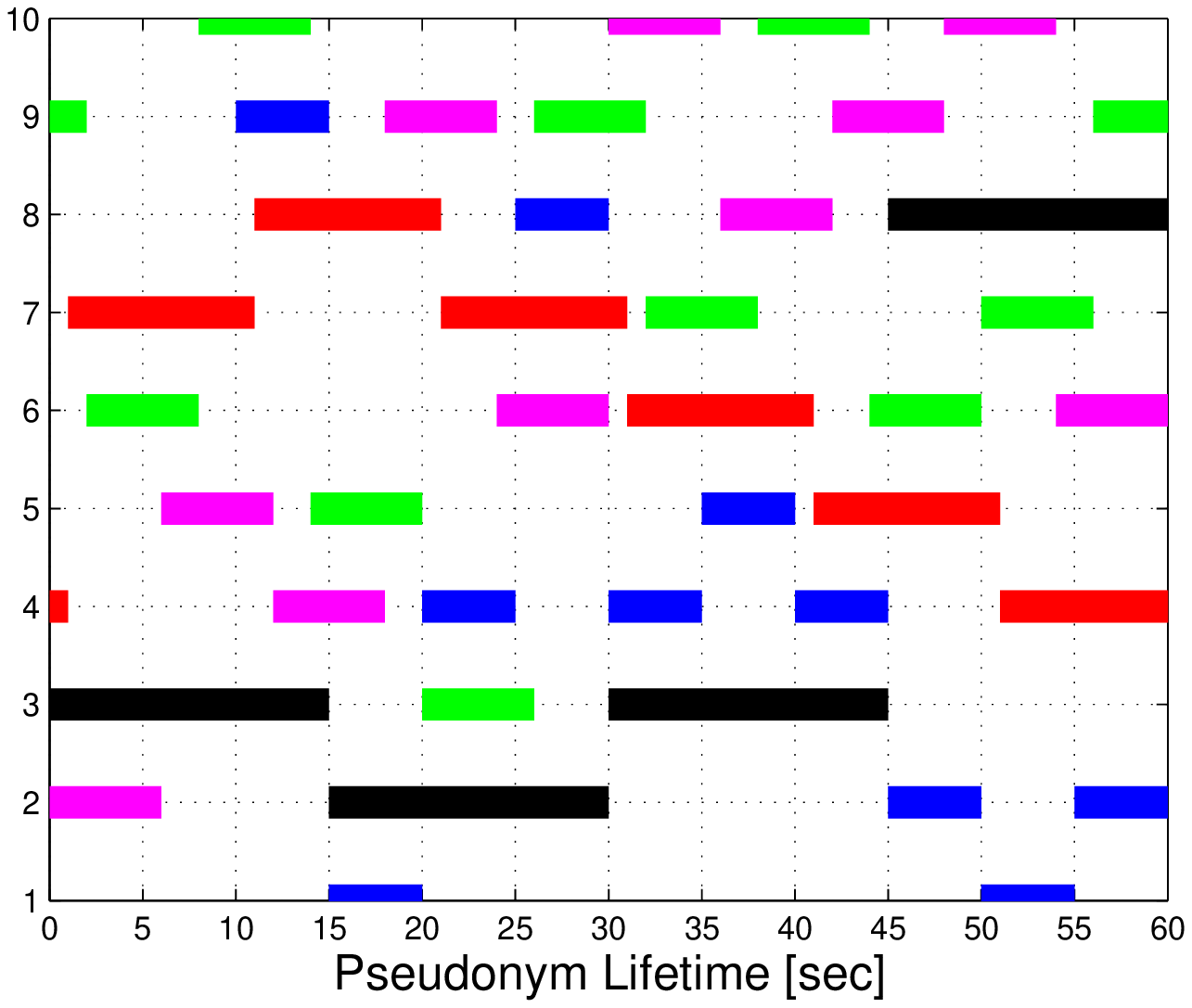}}
      \subfloat[Fixed lifetimes] %Non-overlapping Psnyms
		{\includegraphics[trim=1.5cm 0.1cm 0.7cm 0.65cm, clip=true, totalheight=0.148\textheight,angle=0]{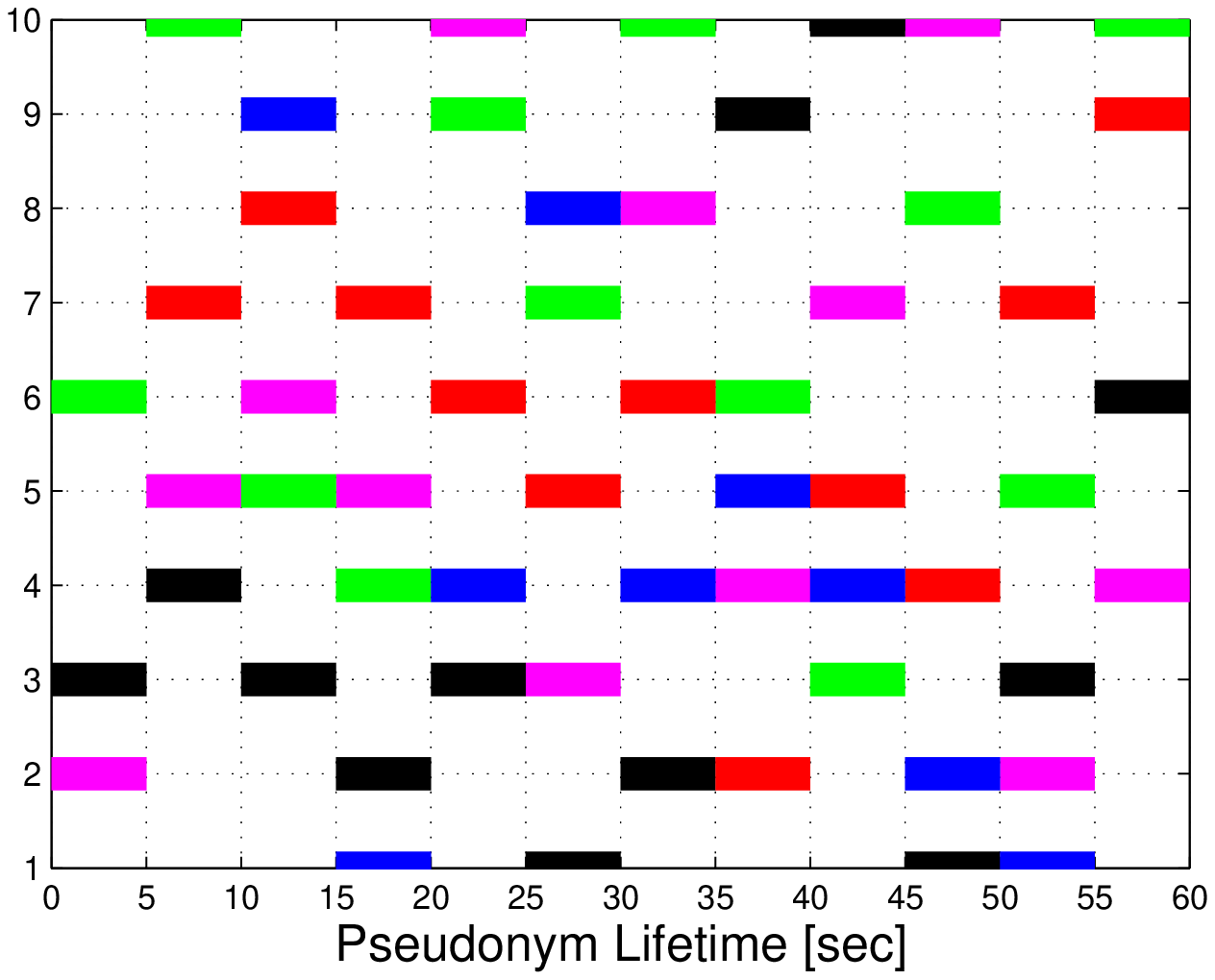}}
      \caption{Pseudonym lifetime policy (Each color shows the non-overlapping pseudonym lifetimes for a vehicle.)}
      \label{fig:linkability-of-psnyms-tkts-using-their-lifetime}
	\vspace{-1em}
\end{figure}

Consider a set of vehicles and their pseudonyms issued in a fully flexible, on-demand manner, with a policy allowing pseudonyms with any lifetime. A transcript of pseudonymously authenticated messages can lead to linked pseudonyms simply by inspecting their (successive) lifetimes. In Fig. \ref{fig:linkability-of-psnyms-tkts-using-their-lifetime}.a, the first pseudonym in row 7 is the only one valid at time 1, with lifetime 10; and then there is a single pseudonym (in row 8, from the same vehicle) valid at time 11, thus making the linking of the two trivial.

In contrast, Fig. \ref{fig:linkability-of-psnyms-tkts-using-their-lifetime}.b illustrates how one can mitigate this vulnerability: the vehicle can request pseudonyms at any point in time, but the \ac{PCA} must, according to the domain policy, issue all its pseudonyms according to fixed lifetimes, equal for all requesting vehicles. With the additional requirement to ensure that there is no backward overlap, by controlling the request interval, the \ac{PCA} aligns all pseudonym lifetimes. This removes any distinction among different sets, resulting in an anonymity set equal to the number of (active) requests. A similar policy should also be applied to ticket acquisition. Although [$t_s$, $t_e$] is protocol selectable, the \ac{LTCA} fixes this to be the same for all tickets; this prevents a \ac{PCA} that serves successive requests from linking tickets (similarly to the flexible pseudonym lifetime case). Of course, the actual request from the vehicle to the \ac{PCA} can be any sub-interval of [$t_s$, $t_e$]. 

Clearly, there is a trade-off in this approach: the longer the interval to obtain pseudonyms (or the ticket validity), the less frequent the vehicle-\ac{VPKI} communication, but the higher the chance to overprovision a vehicle (e.g., if the period includes no movement), and vice-versa. Further discussion on an optimal choice and implementation considerations (e.g., connection to the vehicle mobility, level of unlinkability and thus pseudonym lifetime) are out of the scope of this paper.

%#######################################
%#######################################

%% file: evaluation.tex
\section{Evaluation}
\label{sec:evaluation}

%#######################################
\begin{table*}
	\caption{System Setup and Policies}
	\centering
	 \resizebox{0.85\textwidth}{!}
	 {
	    \begin{tabular}{l | *{5}{c} r}
		      Type of Experiments & Pseudonyms & Execution Time & Frequency & Number of Vehicles \\
		      \hline
		      Ticket \& Pseudonym Acquisition (Fig. \ref{fig:system-performance-for-different-psnym-numbers}) & 1-1K & 50 times & - & - \\
   		      Pseudonym Acquisition (Fig. \ref{fig:LTCA-stress-test-1-ticket-provision} \& \ref{fig:LTCA-PCA-stress-test-obtaining-1-tkt-100-psnyms}.a) & 100 & 1h & 6/h & 10K \\
   		      Pseudonym Acquisition (Fig. \ref{fig:LTCA-PCA-stress-test-obtaining-1-tkt-100-psnyms}.b) & 10-200 & 1h & 10/h & 10K \\
		      \acs{DDoS} Attacks (Fig. \ref{fig:SIS-under-DoS}) & 100 & 1h & 6/h (Attackers: 360/h) & 10K (Attackers: 0-20K) \\
		      Fetching \ac{CRL} (Fig. \ref{fig:CRL-OCSP-under-stress-test}.a) & 1K-100K (Revoked) & 1h & 6/h & 10K \\
		      \ac{OCSP} Operations (Fig. \ref{fig:CRL-OCSP-under-stress-test}.b) & 0.5K-4K (Revoked) & 1h & 6/h & 10K \\
		      Pseudonym Resolution (Fig. \ref{fig:psnym-resolution-performance}) & 1 & 50 times & - & - \\
	    \end{tabular}
	    \label{table:sys-config-policies}
	 } %\end{minipage}}
\end{table*}
%#######################################

We are primarily interested in assessing the efficiency of the full-blown implementation of our \ac{VPKI}, notably measuring the \textbf{performance} both on the client (vehicle) and the server sides. We capture this by measuring \emph{protocol execution delays} experienced on the vehicle side, as well as measuring delays for individual protocol steps. To gauge the \textbf{availability} of the system, its ability to remain operational in the face of failures, we perform two experiments: (i) a \emph{crash} failure of one \ac{PCA}, and (ii) a \ac{DDoS} of increasing intensity against a \ac{PCA} and an \ac{LTCA}.

\textbf{Summary of findings:} We demonstrate a highly efficient \ac{VPKI} system, comparing its performance to the state-of-the-art for the similar experimental setup. In particular, we have a \emph{four-fold} acceleration compared to the best previous results. Essentially, the more efficient the \ac{VPKI} and the lower the overhead/cost for the vehicles, the higher the scalability of the system (being able to service more vehicles per deployed processing power unit). At the same time, the more effective and easier the vehicle-\ac{VPKI} interactions are. Finally, as expected, back-up processing power renders the system dependable.

\subsection{Experimental Setup}\label{subsec:setup}

We allocate resources to distinct \ac{VPKI} servers and we emulate the \ac{VC} system, notably, the population of registered vehicles. We carry out the experiments in a controlled virtualized environment,  with servers and vehicles running on \acp{VM}. This essentially eliminates network propagation delays, which would vary greatly based on the vehicle-\ac{VPKI} connectivity, thus allowing us to isolate the effect of our protocols. Tables \ref{table:sys-config-policies} and \ref{table:vm-SIS-clients-specifications} show the system setup and the servers and clients specifications respectively. We consider large sets of clients, 10K threads on 25 \acp{VM}, executing protocols with (sending requests to) the \ac{VPKI} servers. We experiment under various conditions (configurations, parameters and policies). We gradually increase the load to investigate the behavior of the \ac{VPKI} servers. Our implementation is in C++ and we use OpenSSL for cryptographic operations and algorithms, including: \ac{ECDSA} and \ac{TLS}. As \ac{ETSI} and \acs{IEEE} 1609.2 propose, the \ac{ECDSA} key size is 256 bits \cite{ETSI-102-638, 1609draft}, although other key sizes are also acceptable. It is important to note that the emulated vehicle resources are modest, much lower than the anticipated cryptographic processing power in ongoing \acp{FOT} (e.g., \cite{preserve-url}). Thus, one can expect even better results overall. 

%#######################################
\begin{table}
	\caption{Servers and Clients Specifications}
	\centering
	  \resizebox{0.4\textwidth}{!}
	  {
	    \begin{tabular}{l | *{4}{c} r}
		       & \ac{LTCA} & \ac{PCA} & \ac{RA} & Clients \\
		      \hline
		      \acs{VM} Number & 2 & 5 & 1 & 25 \\
		      Dual-core CPU (Ghz) & 2.0 & 2.0 & 2.0 & 2.0 \\
		      BogoMips          & 4000 & 4000 & 4000 & 4000 \\
		      Memory            & 2GB & 2GB & 1GB & 1GB \\
		      Database          & MySQL & MySQL & MySQL & MySQL \\
		      Web Server        & Apache & Apache & Apache & - \\
		      Load Balancer		& Apache & Apache & - & - \\
		      Emulated Threads  & - & - & - & 400
	    \end{tabular}
	    \label{table:vm-SIS-clients-specifications}
       }
\end{table}

\subsection{Results \& Analysis}\label{subsec:results-and-analysis}

\subsubsection{Ticket \& Pseudonym Provisioning}\label{subsubsec:ticket-pseudonym-acquisiyion}

\begin{figure}[t!]
	\centering
	\begin{minipage}{0.25\textwidth}
		  \centering
		  \includegraphics[width=\textwidth,height=\textheight,keepaspectratio]{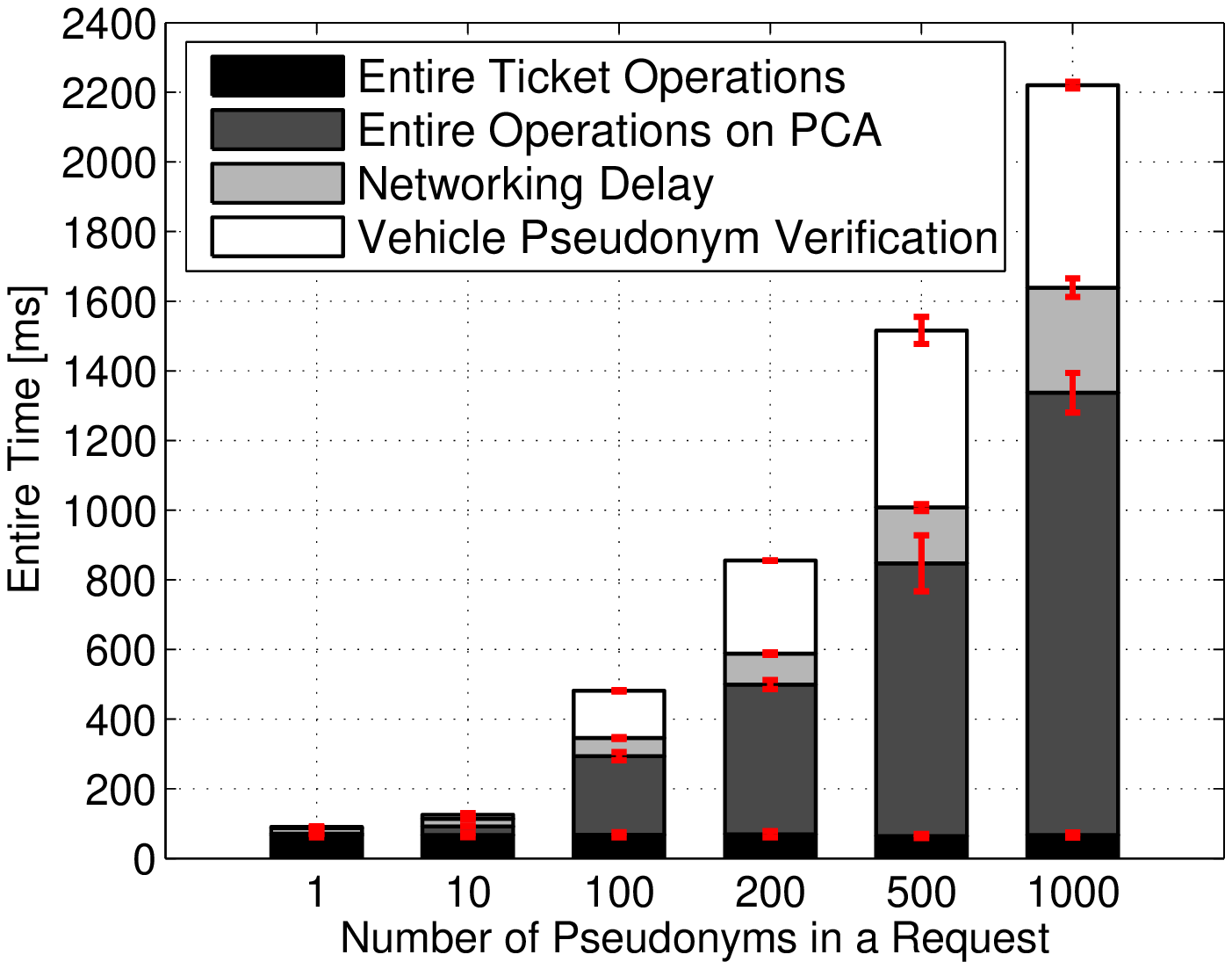}
		  \caption{Client processing time}
		  \label{fig:system-performance-for-different-psnym-numbers}
	\end{minipage}%
	\begin{minipage}{0.25\textwidth}
		  \centering
		  \includegraphics[width=\textwidth,height=\textheight,keepaspectratio]{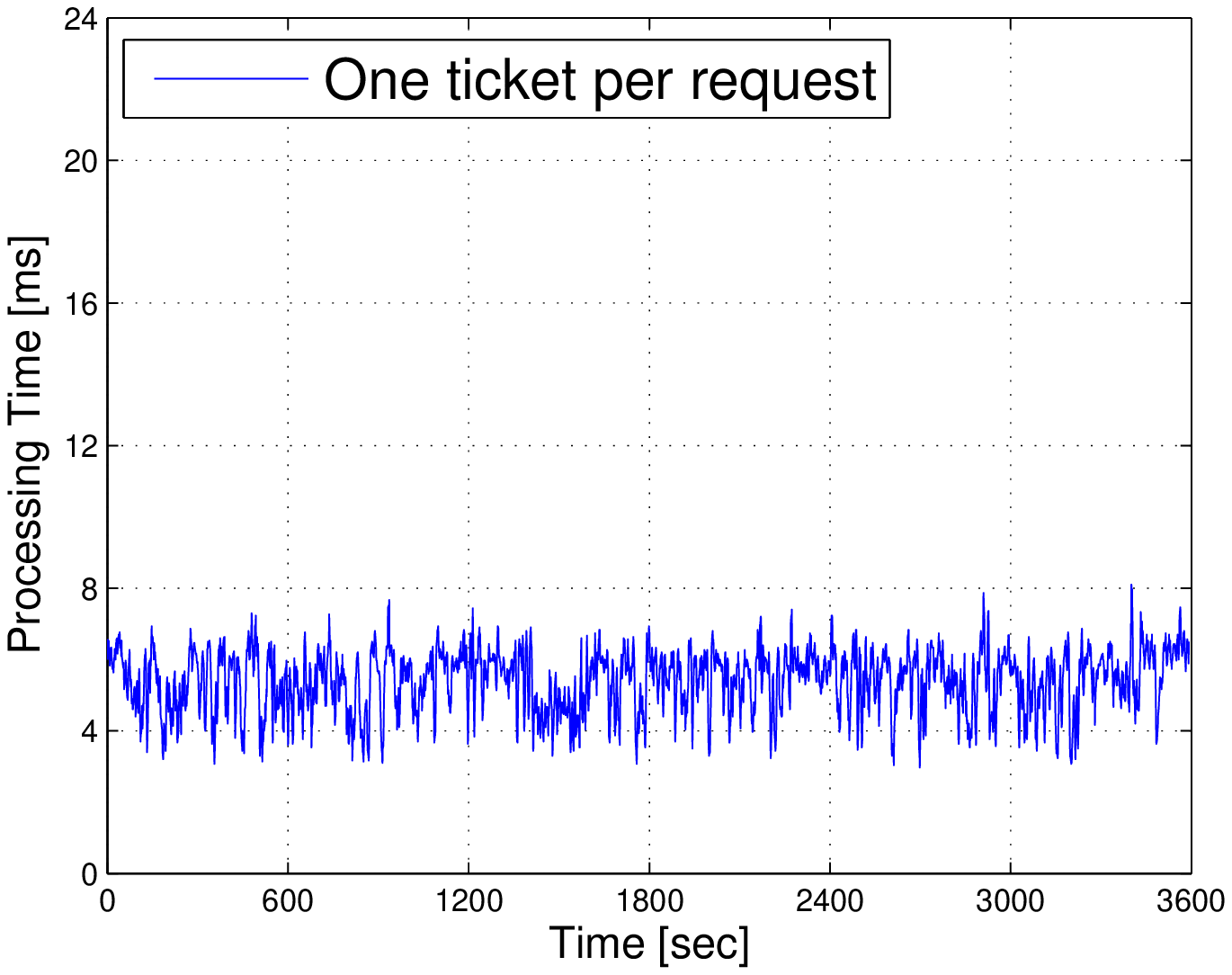}
		  \caption{\ac{LTCA} performance}
		  \label{fig:LTCA-stress-test-1-ticket-provision}
	\end{minipage}
\end{figure}

Fig.~\ref{fig:system-performance-for-different-psnym-numbers} shows the delay to obtain pseudonyms for each component, ticket provisioning, pseudonym verification, the \ac{PCA} processing delay and network transmission delay.\footnote{Again, the lab environment dwarfs this, but it is deliberate to factor this out as it is orthogonal to our design. It was also done so in the works we compare to. A separate investigation taking into account the access networks is interesting future work.} The processing time to generate public/private keys is not considered here, as they can be generated off-line on the vehicle. For example, the delay to obtain 100 pseudonyms is around 500 ms. Fig. \ref{fig:LTCA-stress-test-1-ticket-provision} shows the average response time for the \ac{LTCA} to issue a ticket, approximately 5 ms (including request decapsulation, verification of the \ac{LTC}, ticket issuance and response encapsulation).

%#######################################
%#######################################

%#######################################
\begin{figure}[t!]
	\captionsetup[subfigure]{labelformat=empty}
      \subfloat[(a) Issuing 100 pseudonyms per request]{
		\includegraphics[trim=1.7cm 0cm 1.5cm 1.1cm, clip=true, totalheight=0.145\textheight,angle=0]{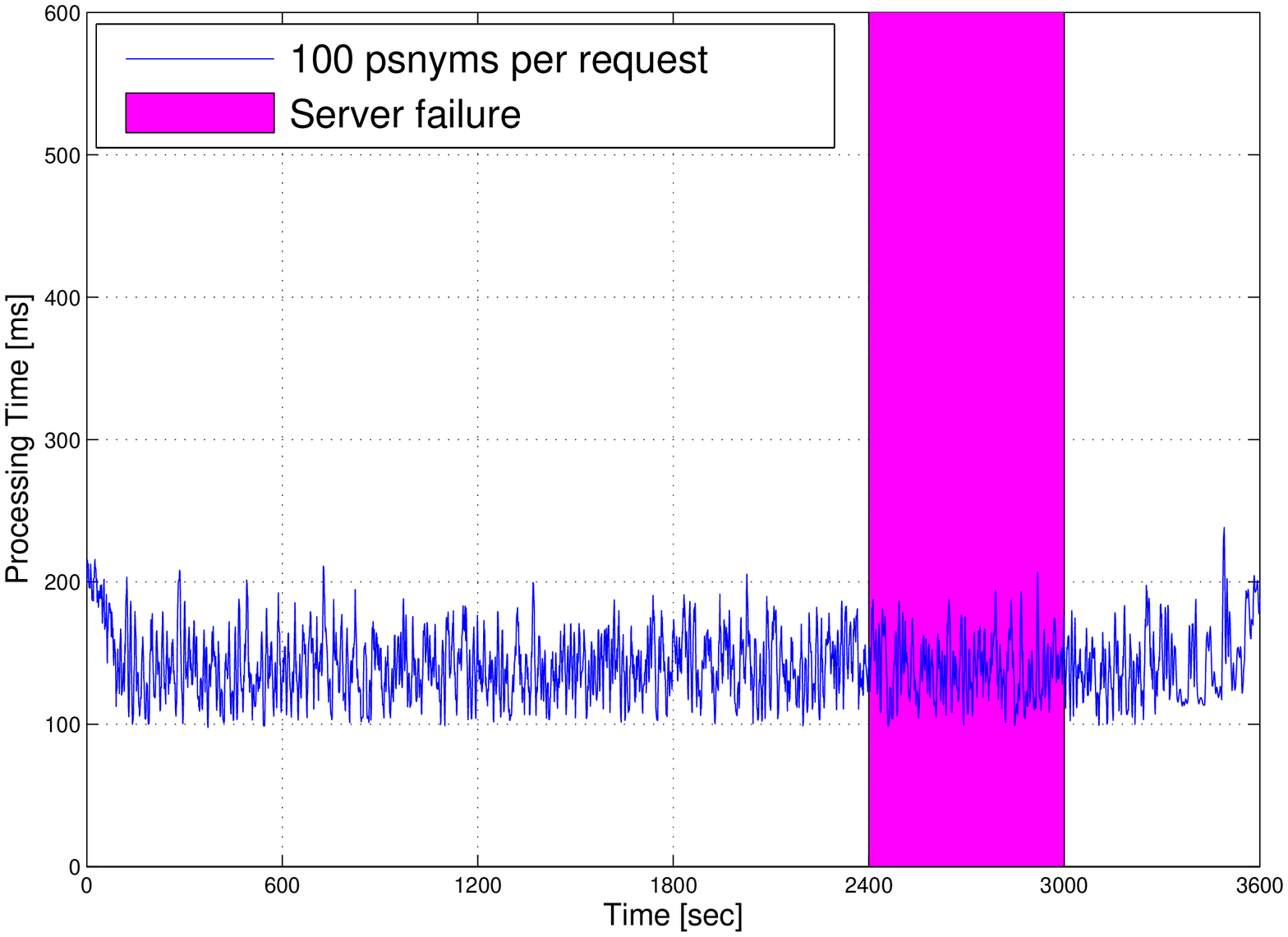}}
        \subfloat[\hspace{4mm}(b) Performance under different \newline \hspace*{15mm} configurations]
		{\includegraphics[trim=0.6cm 0cm 0cm 0.689cm, clip=true, totalheight=0.145\textheight,angle=0]{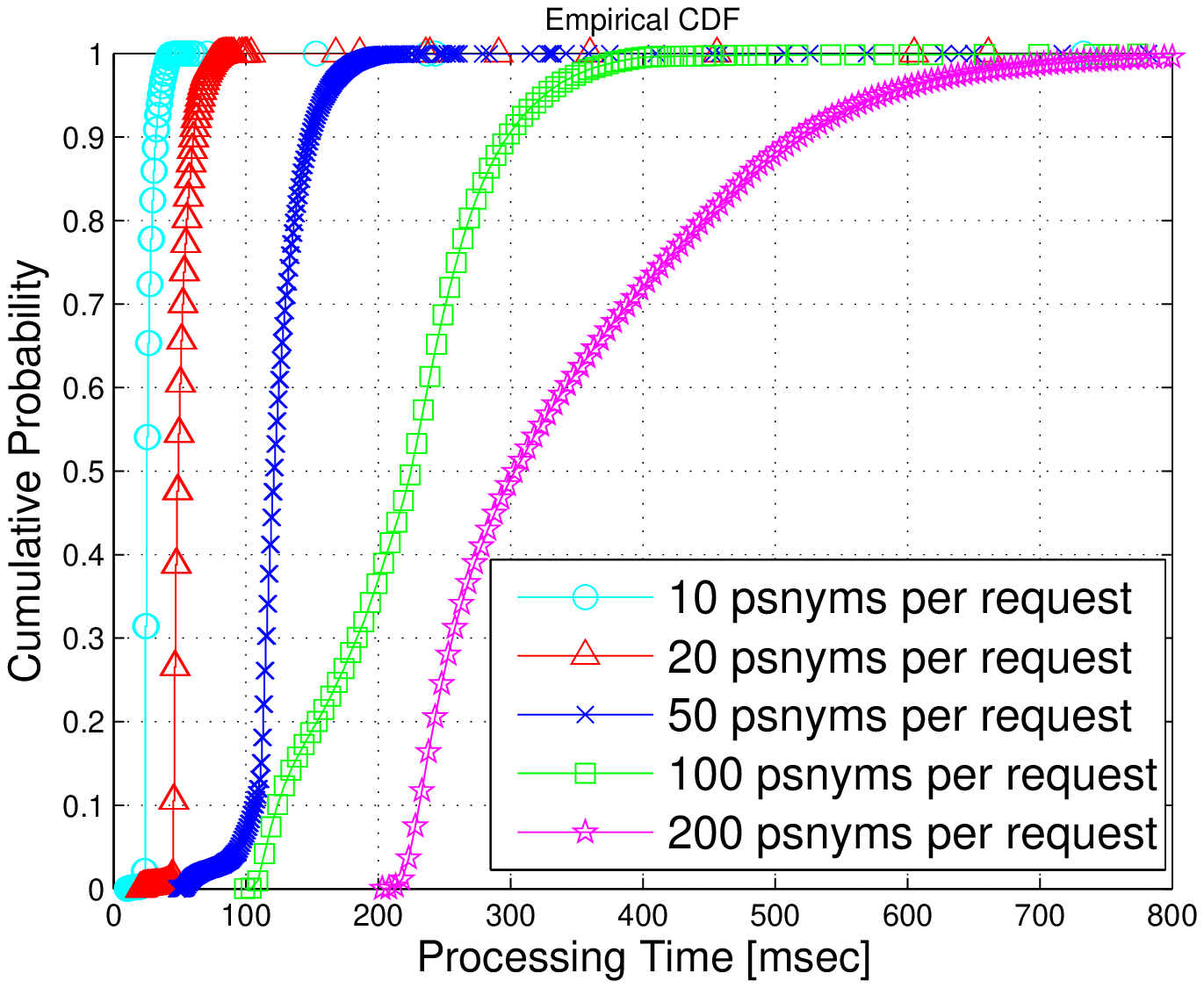}}
      \caption{\ac{PCA} processing delay}
      \label{fig:LTCA-PCA-stress-test-obtaining-1-tkt-100-psnyms}
	\vspace{-1em}
\end{figure}
%#######################################

Fig. \ref{fig:LTCA-PCA-stress-test-obtaining-1-tkt-100-psnyms}.a shows the average response time of the \acp{PCA} for issuing 100 pseudonyms (request decapsulation, ticket certificate chain verification, verification of proof of possessions for each public key, issuance of pseudonyms and encapsulation of the response). The system configuration here requests from the \acp{PCA} to issue 100 pseudonyms every 100 minutes per vehicle. The shadowed area shows the period during which one of the servers instantiating the \acp{PCA} was forced to crash. We see that the responsiveness of the \ac{PCA} remains unscathed.

Fig. \ref{fig:LTCA-PCA-stress-test-obtaining-1-tkt-100-psnyms}.b shows the performance of the \acp{PCA} issuing different numbers of pseudonyms. For example, the cumulative probability of delays for obtaining 200 pseudonyms is: $F_x(t=500)=0.9$, or $Pr\{t\leq500\}=0.9$. This confirms the system can scale: obtaining more than 120 pseudonyms every 10 minutes (pseudonym lifetime of 5 sec.) is considered as a quite demanding case, considering current expectations that one pseudonym per day or per trip is used \cite{c2c}. Of course, this is a function of the number of registered vehicles and the \ac{PCA} processing power. But we see that a modest machine can serve thousands of demanding vehicles. 

\begin{figure} [t!]
  \begin{center}
    \centering
      \subfloat[\ac{LTCA} performance]{
		\includegraphics[trim=0.8cm 0.25cm 0.7cm 0.63cm, clip=true, totalheight=0.145\textheight,angle=0]{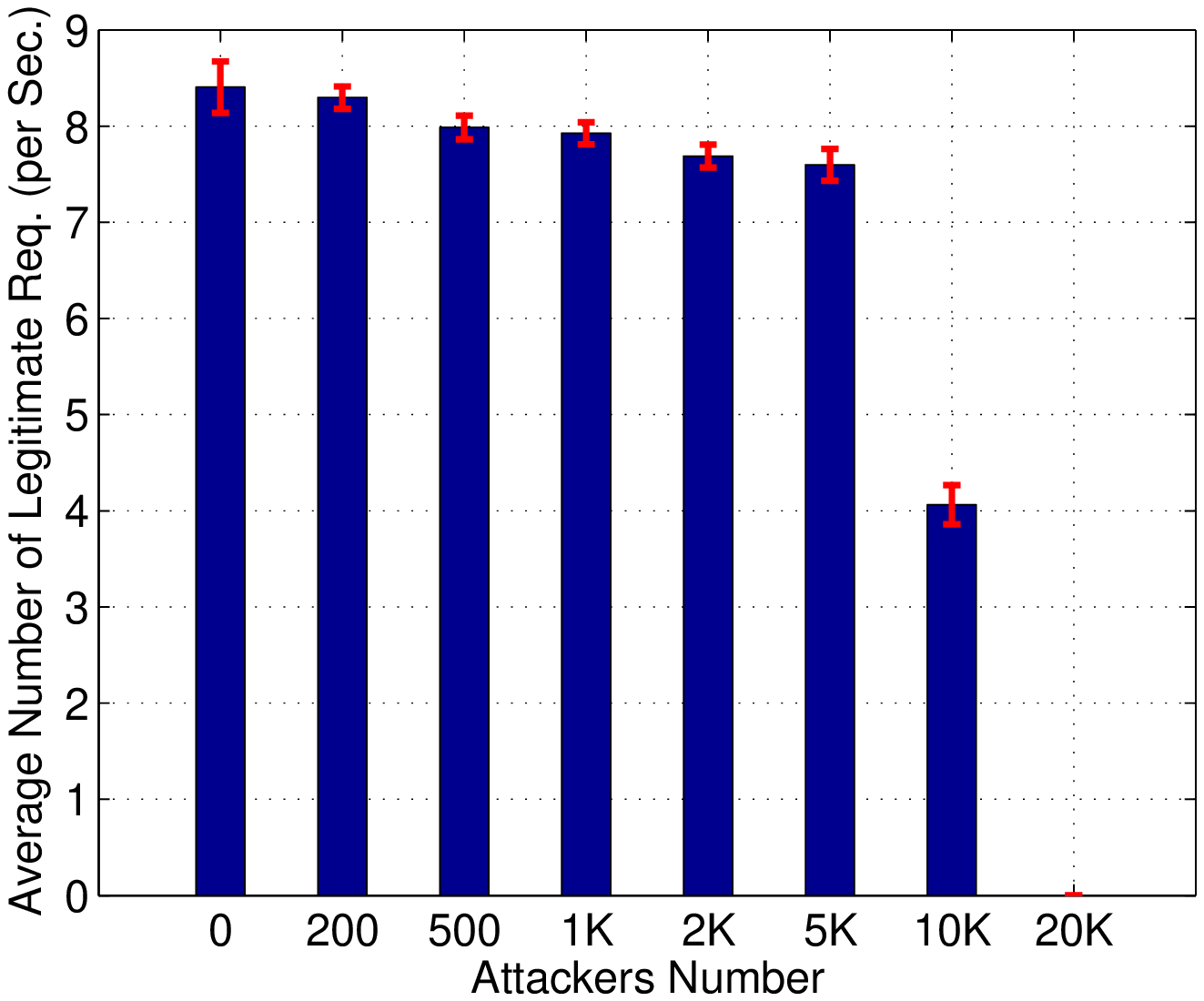}}
      \subfloat[\ac{PCA} performance]
		{\includegraphics[trim=0.5cm 0.25cm 0.7cm 0.63cm, clip=true, totalheight=0.145\textheight,angle=0]{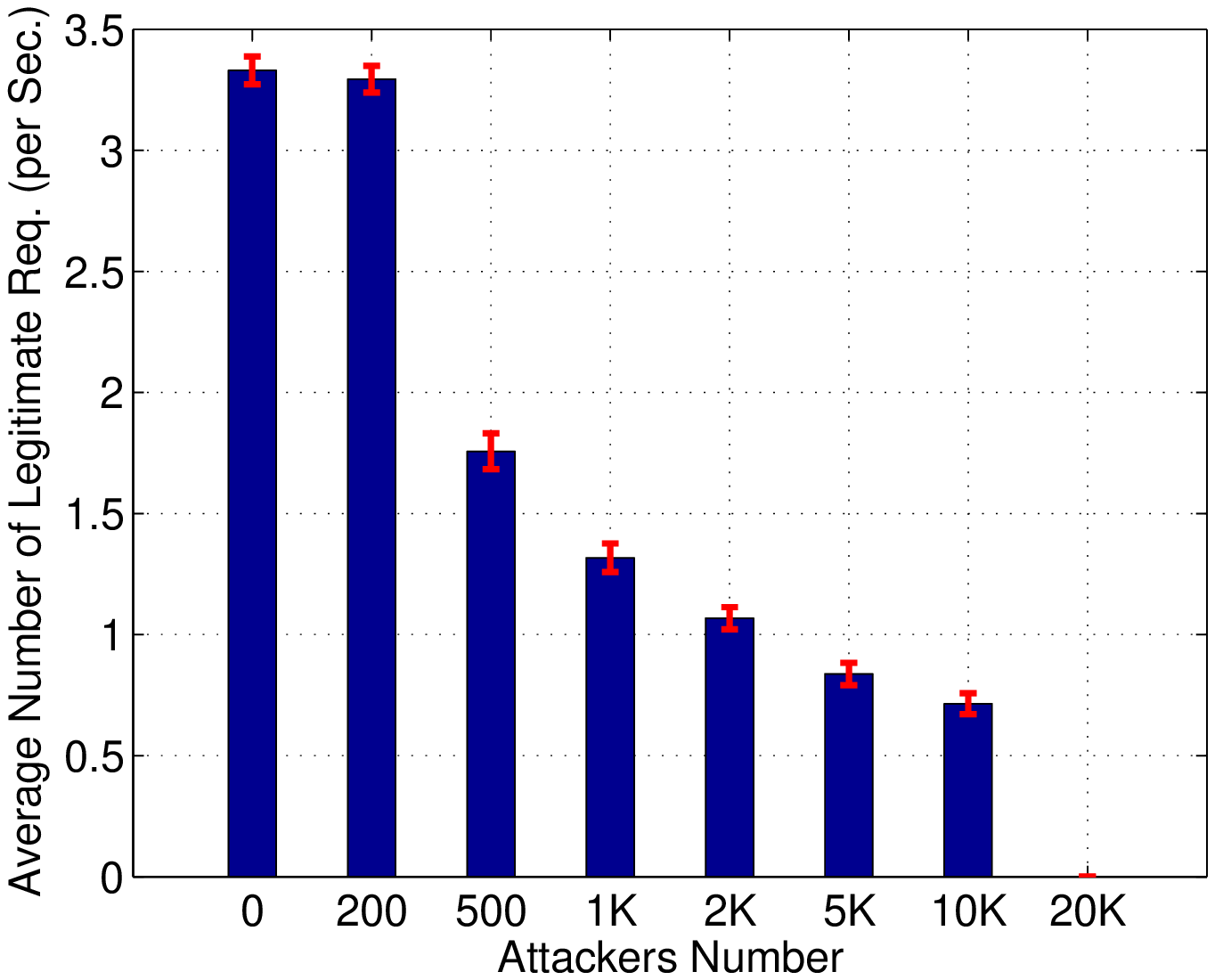}}
      \caption{\ac{VPKI} servers under a \ac{DDoS} attack}
      \label{fig:SIS-under-DoS}
	\vspace{-1em}
  \end{center}
\end{figure}

\textbf{\ac{DDoS} Attack:} External adversaries could clog the \ac{LTCA} or the \ac{PCA} with spurious, bogus requests.\footnote{Similarly for the \ac{RA}, but as it is less critical for the real-time operations of the \ac{VC} system, we do not investigate here.} They can send requests to the \ac{LTCA} with fake certificates, or to the \ac{PCA} with fake tickets. We set a high request frequency, on the average once per 10 seconds, increasing the number of adversarial nodes acting this way up to 20K. Fig.~\ref{fig:SIS-under-DoS} shows the responsiveness of an \ac{LTCA} and a \ac{PCA} under \ac{DDoS}. The average number of legitimate serviced requests per second for the \ac{LTCA} drops by half for 10K attackers. While the same happens for the \ac{PCA} only for 500 attackers. This is naturally so, because both servers have the same resources, but the overhead for providing pseudonyms is much higher than that for providing a ticket. One can suggest allocating higher resources to the \ac{PCA}, or employing \ac{DDoS} mitigation techniques appropriate for the limited client (vehicle) resources, e.g., puzzles \cite{aura2001resistant}.

%#######################################
%#######################################

%*******************************************************************************
\subsubsection{Revocation}
\label{subsubsec:revocation-performance}

\begin{figure}[t!]
  \begin{center}
    \centering
      \subfloat[Obtaining a \ac{CRL}]{
 		\includegraphics[trim=0.75cm 0.2cm 1.1cm 0.7cm, clip=true, totalheight=0.145\textheight,angle=0]{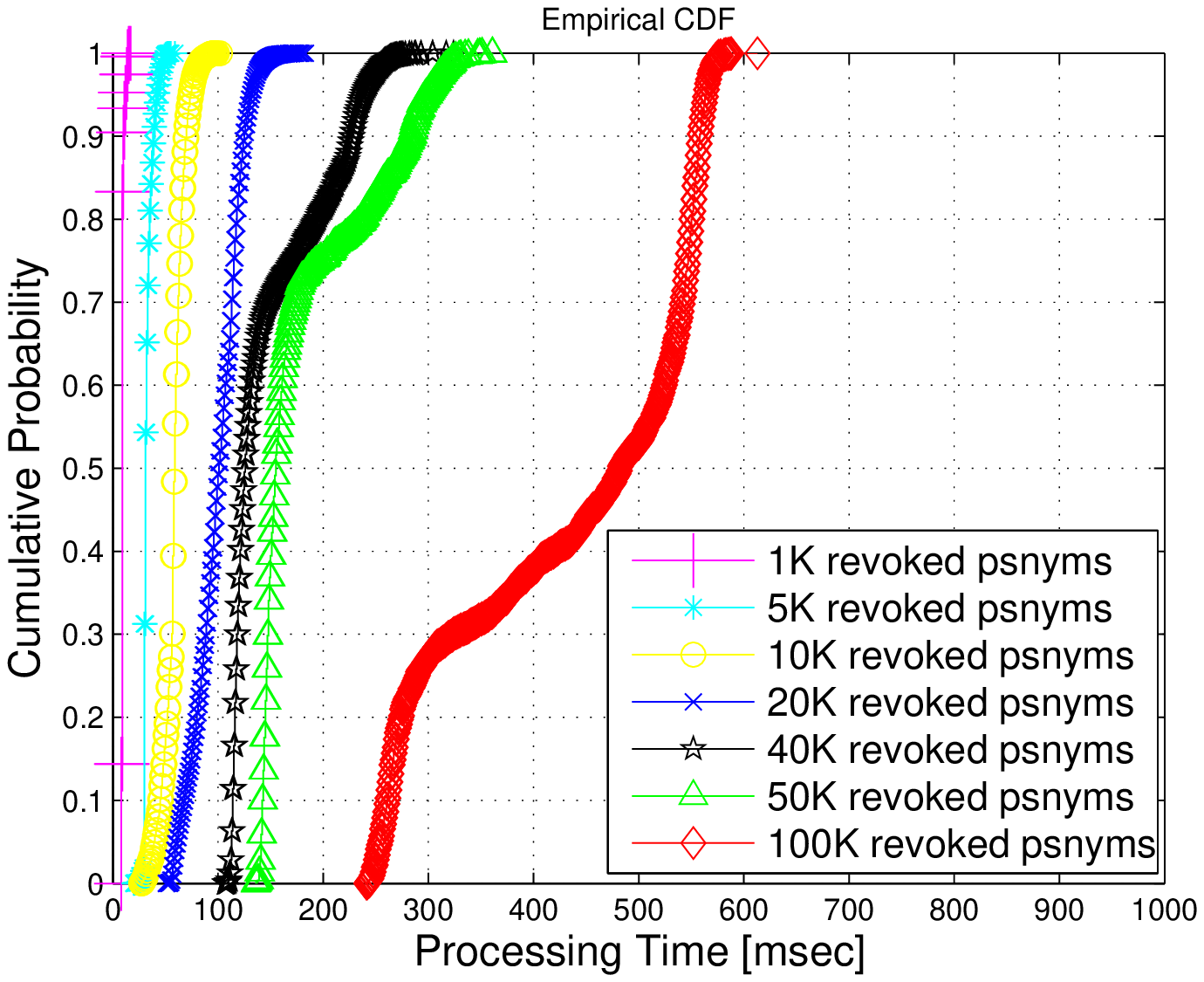}}
       \subfloat[\ac{OCSP} validation]
		{\includegraphics[trim=0.5cm 0.2cm 1.2cm 0.7cm, clip=true, totalheight=0.145\textheight,angle=0]{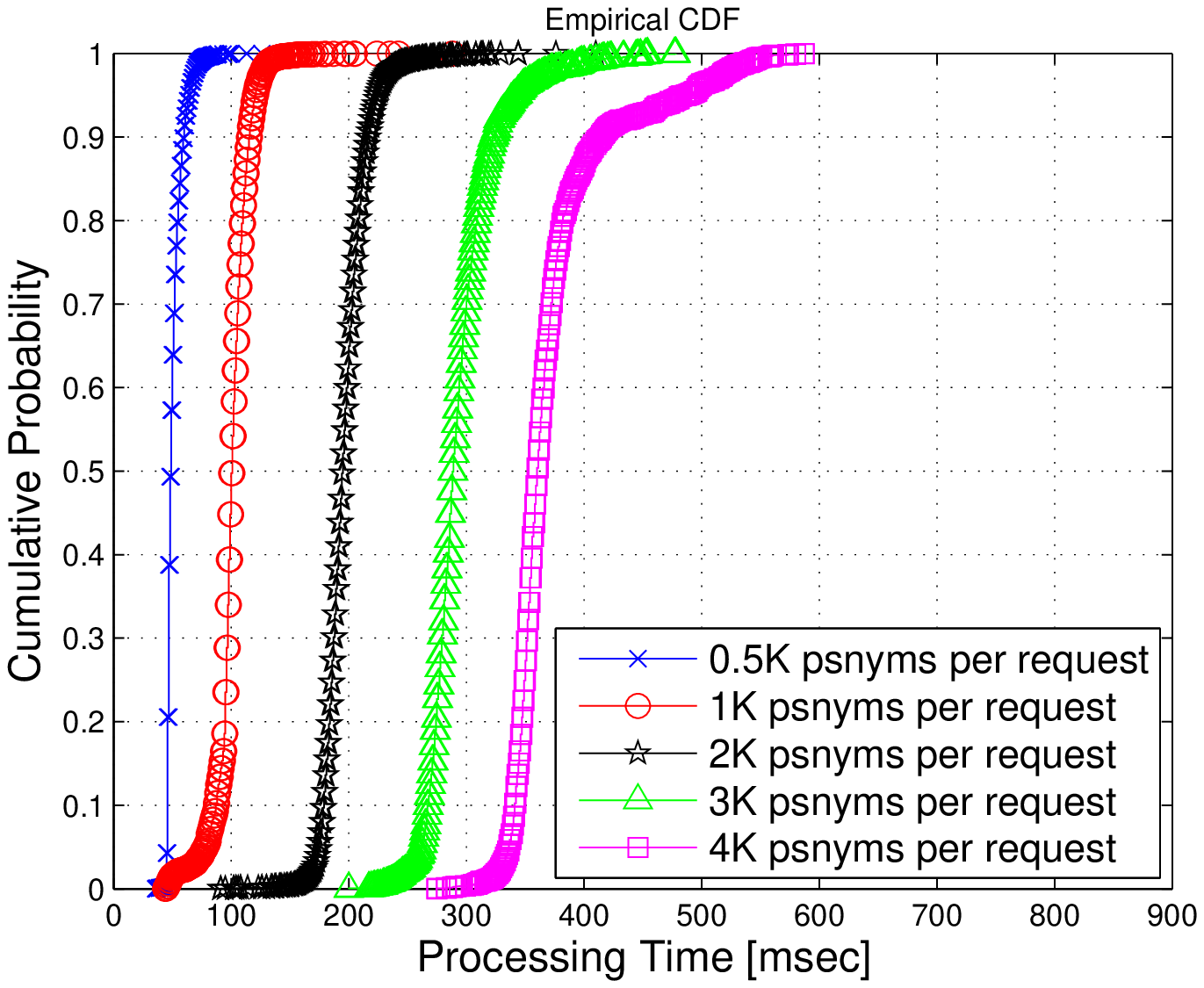}}
      \caption{Obtaining a \ac{CRL} or performing \ac{OCSP} validation}
      \label{fig:CRL-OCSP-under-stress-test}
	\vspace{-1em}
  \end{center}
\end{figure}

Fig. \ref{fig:CRL-OCSP-under-stress-test}.a illustrates performance when 10K vehicles query the \acp{PCA}, once every 10 minutes on average, to fetch the \ac{CRL}; with different numbers of revoked pseudonyms, from one to one-hundred thousand (1K to 100K). For example, for a \ac{CRL} with 50K revoked pseudonyms, $F_x(t=280)=0.9$, or $Pr\{t\leq280\}=0.9$. Fig. \ref{fig:CRL-OCSP-under-stress-test}.b shows the performance when the same client population checks the revocation status of 500 to 4,000 different pseudonyms. We reckon that \ac{OCSP} would not be used for such high numbers of pseudonyms. Still, the system can comfortably handle such demanding load.

\subsubsection{Pseudonym Resolution}
\label{subsubsec:pseudonym-resolution-performance}

\begin{figure} [t!]
    \centering
	\includegraphics[trim=3cm 0.7cm 3cm 1cm, clip=true, totalheight=0.19\textheight,angle=0]{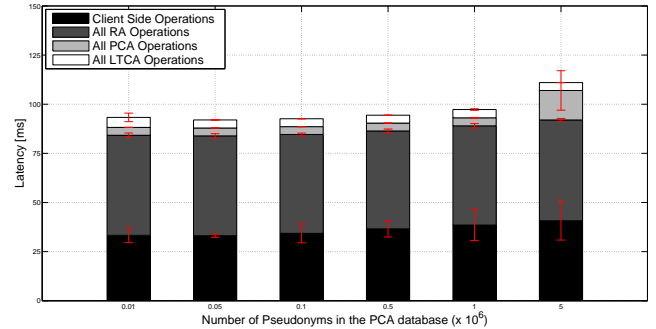} % picture file name
    \caption{Response Time to Resolve \& Revoke a Pseudonym}
    \label{fig:psnym-resolution-performance}
	\vspace{-1em}
\end{figure}

Fig. \ref{fig:psnym-resolution-performance} shows the delay to resolve and revoke a pseudonym. As \ac{PCA} databases could be gigantic, we evaluate resolution for 10,000 to 5 million pseudonyms. Pseudonym resolution and revocation is comfortably handled, in around 100 ms.

\subsubsection{Performance Comparison}
\label{subsubsec:results-comparison}

We compared the relevant results\footnote{\ac{CRL} and \ac{OCSP} operations and resiliency to \ac{DDoS} attacks were not considered in related works.} directly to the performance results presented in \cite{vespa, alexiou2013dspan, gisdakis2013serosa}, exactly because we use very similar setup. We see we achieve significant improvements in terms of efficiency and performance. The main reasons for such significant improvements, given protocols of very similar message complexity, are: multi-threading implementation and use of database, code and memory usage optimization techniques. The result of these can be a 4-fold improvement, e.g., the processing delay to issue 10 pseudonyms for the \ac{PCA} for SEROSA \cite{gisdakis2013serosa} is around 100 ms, while it is approx. 26 ms in our system.

%#######################################
%#######################################

%#######################################
%#######################################

%% file: related-works.tex
\section{Related Work}
\label{sec:related-work}

The \ac{SeVeCom} project \cite{SeVeCom}, and its continuation, \ac{PRESERVE} \cite{preserve}, have led to a \ac{VPKI} instantiation compliant to the \ac{C2C-CC} framework. Because of direct \ac{PCA} - \ac{LTCA} communication at the time of pseudonym provision, the \ac{LTCA} knows the pseudonym providing \ac{PCA}, thus it can easily link messages. Similarly, the \ac{SCMS} \cite{whyte2013security} requires that the identity provider forwards requests to \acp{PCA}, thus being prone to the same inference.\footnote{Unlike the PRESERVE system, \ac{SCMS} allows multiple simultaneously valid pseudonyms held by the vehicle, thus not being concerned with Sybil-based misbehavior.} The linking of the pseudonym request (and thus long-term identity) to a specific \ac{PCA} and the request timing (and thus an easy to guess set of pseudonyms and signed messages) is possible for VeSPA and its extension \cite{vespa,alexiou2013dspan}: they leverage an anonymized ticket but the \ac{LTCA} still knows \emph{when} the ticket will be used to obtain pseudonyms and from \emph{which} \acp{PCA}. Our \ac{VPKI} addresses this concern, providing improved security. We also achieve significant improvement in performance. For example, to obtain 10 pseudonyms, the processing delays of the \acp{PCA} in VeSPA and our scheme are 300 ms and 26 ms, respectively. 

Concerning pseudonym resolution, the V-token scheme \cite{schaub2010v} mandates a random checking of V-tokens to prevent vehicles from binding false identities to pseudonyms. However, an honest-but-curious \ac{LTCA} would learn which vehicles the randomly checked V-tokens belong to. Having this information, it can link the (eavesdropped) pseudonyms to the vehicle identities by looking at the V-tokens included in the pseudonyms. CoPRA \cite{bibmeyer2013copra} stores resolution IDs, the hash of each short-term public key and a random number, at both the \ac{LTCA} and the \ac{PCA}; the honest-but-curious \ac{LTCA} can easily calculate hashes of any public key from (eavesdropped) pseudonyms and link them to the identity of the vehicle. Our proposal is more general and more robust, further analyzed, and with detailed experimental evaluation. 

SEROSA \cite{gisdakis2013serosa} proposed a general service-oriented security architecture seeking to bridge Internet and the \ac{VC} domains. However, the identity provider can still infer the identity of the service provider based on the protocol design. Moreover, the multi-domain environment explicitly addressed by SEROSA (as was the case for VeSPA, and not elaborated in other proposals) leaves space for Sybil-based misbehavior. The infrastructure cannot prevent multiple spurious requests to different \acp{PCA}. Of course, an \ac{HSM} (ensuring all signatures are generated under a single valid pseudonym at any time) can be a general remedy to the problem \cite{papadimitratos2007architecture}. Our \ac{VPKI} alone prevents Sybil-based misbehavior without trusted hardware and it is significantly more efficient in terms of performance.

Last but not least, an issue relevant to all systems with non-overlapping pseudonym lifetimes: while seeking to prevent Sybil-based misbehavior even without trusted hardware (as long as the misbehaving client cannot requests to multiple \acp{PCA} serviced), linkability  could remain easy (Sec. \ref{sec:security-analysis}). This was not identified before and our \ac{VPKI} hints how to address it through its pseudonym lifetime and ticket validity policy.

Mixing anonymous authentication with classic public key cryptography was proposed: e.g., in \cite{PapadiCLH:C:08} group signatures allow pseudonym self-generation on-the-fly; or, in \cite{studer2009tacking} group keys are used as long-term credentials instead of certificates. Standardization bodies have not yet embraced such approaches. Moreover, performance evaluation results are available for \ac{V2V} communication \cite{PapadiCLH:C:08,calandriello2011performance}, but not for a full-blown credential management system.

%% file: conclusions.tex
\section{Conclusions}
\label{sec:conclusions}

Our results primarily show that our \ac{VPKI} strengthens security and privacy protection, extends functionality, and outperforms (in response delays) prior proposals. These contributions lead us closer to a robust and scalable \ac{VPKI}.